\documentclass[prl,aps,nofootinbib,superscriptaddress,twocolumn,showpacs]{revtex4-1}

\usepackage{graphicx}
\usepackage{epsfig}
\usepackage{amsmath}
\usepackage{amsfonts}
\usepackage{bm}
\usepackage{xcolor}
\usepackage{appendix}
\usepackage{verbatim} 
\usepackage{enumitem} 
\usepackage{dcolumn} 

\def\ShowChangesVersion{n}        

\newcommand{\bea}{\begin{eqnarray}}
\newcommand{\eea}{\end{eqnarray}}
\newcommand{\beq}{\begin{equation}}
\newcommand{\eeq}{\end{equation}}
\newcommand{\bqa}{\begin{eqnarray}}
\newcommand{\eqa}{\end{eqnarray}}

\def\mqo2{{\!\!\!}}

\newcommand{\etal}{{\it et al.{}}}
\newcommand{\HBCPT}{HB$\chi$PT}
\newcommand{\alphae}{\ensuremath{\alpha_{E1}}}
\newcommand{\betam}{\ensuremath{\beta_{M1}}}
\newcommand{\fm}{\ensuremath{\mathrm{fm}}}
\newcommand{\chiEFT}{$\chi$EFT}
\newcommand{\mpi}{m_\pi}
\newcommand{\MN}{\ensuremath{M_\mathrm{N}}} 

\def\AnswerYes{y}   
\ifx\ShowChangesVersion\AnswerYes
  \usepackage{ulem}
   
  \newcommand{\add}[1]{\textbf{#1}}               
  \newcommand{\delete}[1]{\sout{#1}}            
  \renewcommand{\emph}[1]{\textit{#1}}           
\else

  \newcommand{\sout}[1]{}
  \newcommand{\xout}[1]{}
  
  \newcommand{\add}[1]{#1}
  \newcommand{\delete}[1]{}
\fi

\begin{document}

\title{Proton polarisabilities from Compton data using Covariant Chiral EFT}
\author{Vadim Lensky}\email{Vadim.Lenskiy@manchester.ac.uk}
\affiliation{Theoretical Physics Group, School of Physics and Astronomy, University of Manchester, Manchester, M13 9PL, United Kingdom}
\affiliation{Institute for Theoretical and Experimental Physics, B.~Cheremushkinskaya 25, 117218 Moscow, Russia}
\author{Judith McGovern}\email{Judith.McGovern@manchester.ac.uk}
\affiliation{Theoretical Physics Group, School of Physics and Astronomy, University of Manchester, Manchester, M13 9PL, United Kingdom}

\date{\today}

\begin{abstract}
We present a fit of the spin-independent electromagnetic polarisabilities of the proton to low-energy Compton scattering data in the framework of 
covariant baryon chiral effective field theory.  Using the Baldin sum rule to constrain their sum, we obtain 
$\alphae=[10.6\pm0.25(\text{stat})\pm0.2(\text{Baldin})
  \pm0.4(\text{theory})]\times10^{-4}\;\fm^3$
  and $\betam =[3.2\mp0.25(\text{stat})\pm0.2(\text{Baldin})
  \mp0.4(\text{theory})]\times10^{-4}\;\fm^3$, in excellent agreement with other chiral extractions of the same quantities.
\end{abstract}
\maketitle

The electromagnetic polarisabilities of the proton have been a subject of investigation for many years; the earliest extractions from low-energy Compton 
scattering data were carried out in the 1950s, and the relevant database was greatly expanded in the 1990s.   In the very low-energy regime one can make an 
expansion of the cross section which 
deviates from the Thomson scattering only through the inclusion of the two spin-independent polarisabilities $\alphae$ and $\betam$.   However, 
since very few measurements have been taken below 80~MeV, almost all extractions require theoretical input to describe the evolution of the cross section with 
energy. Historically, dispersion relation (DR) approaches were used, with input from pion photoproduction data. A model-independent constraint can be obtained 
from the Baldin sum rule, most recently evaluated to give $\alphae+\betam=13.8\pm0.4$ in units of $10^{-4}\;\fm^3$ \cite{Olmos:2001}, so typically the parameter 
extracted is $\alphae-\betam$. In 2001 Olmos de Le\'on \etal\  published the most comprehensive data set yet, obtained with the TAPS detector at MAMI, and 
in a DR framework analysed it together with other ``modern" data to give $\alphae-\betam=10.5\pm0.9\pm0.7$ in the same units \cite{Olmos:2001}. For some time this was 
regarded as the definitive result.
 
However, chiral effective field theories (\chiEFT) can also be used to describe Compton scattering amplitudes.  These are field theories in which the 
interactions of low-energy degrees of freedom are governed by the symmetries of QCD, and scattering amplitudes can be systematically expanded in powers
of the ratio of light to heavy scales. The former are typically external particle momenta of the order of the pion mass, and the latter are governed by those 
particles such as the $\rho$ meson which are not included explicitly in the theory but whose effects, along with other short distance physics, are encoded
in low energy constants. At leading one-loop order in the theory with pion and nucleons these predictions are parameter-free, but beyond leading order 
$\alphae$ and $\betam$ are free parameters which can be fit to  data. The first attempt to do this was the work of Beane \etal\ \cite{Beane:2002wn,Beane:2004ra}, 
working in  heavy baryon (HB) chiral perturbation theory.  The absence of a dynamical Delta isobar restricted the fit to relatively low momentum transfer, 
and as a result the statistical errors were large.  However, the inclusion of the Delta followed shortly~\cite{Hemmert:1996rw,Pascalutsa:2002pi,Hildebrandt:2003fm}.  
Most recently, the result  
$\alphae-\betam=7.5\pm0.7\pm0.6$ has been obtained by McGovern \etal\ in ref.~\cite{McGovern:2012ew}. Although this value is compatible with the previous chiral 
extractions~\cite{Beane:2002wn,Beane:2004ra,Hildebrandt:2003fm}, the calculation was carried out to a sufficiently high order, 
and fit a sufficiently large set of experimental data, that the results were 
precise enough to demonstrate a tension with DR-based results at, roughly, the  2$\sigma$ level (combining all errors in quadrature).\footnote{Because 
the paper is not easily available, the rather better agreement with the work of Baranov \etal\ \cite{Baranov:2001} which, using a DR-based fit to 
world data, finds $\alphae-\betam=9.5\pm1.0\pm0.7$, has been less noted. In fact for ``modern" pre-TAPS data alone, Baranov's result is $7.7\pm 1.2$.}

In addition to $\alphae$ and $\betam$ there are also four spin polarisabilities of the proton.  One combination of these,  $\gamma_0$, satisfies a Baldin-like 
sum rule and so is reasonably well known:  $\gamma_0=-0.90 \pm 0.08 {\rm (stat)} \pm 0.11 {\rm(sys)}$ in units of $10^{-4}\fm^4$, \cite{Pasquini:2010zr}.
In looking for sources of the discrepancy in the extracted values of $\alphae-\betam$, it has been suggested that the fact that the EFT and DR values of 
the spin polarisabilities are quite discrepant is the problem; since those obtained at this order in the EFT are not in good 
agreement with the sum-rule determination of $\gamma_0$, this has 
been used to suggest that the EFT extraction is less reliable, or at least that the errors are underestimated.

In this paper we consider the situation in a different variant of \chiEFT, namely one which does not use the heavy-baryon expansion but treats the nucleon 
fields as Dirac spinors \cite{Gasser:1987rb}.  
Compton scattering amplitudes were first calculated in this approach by Lensky and Pascalutsa in ref.~\cite{Lensky:2009uv}, using (a modification of) the 
EOMS remormalisation scheme \cite{Fuchs:2003qc}.
The power-counting scheme we use is the so-called ``$\delta$-counting", 
with the small parameter $\delta\sim\mpi/\Delta\sim\Delta/\Lambda_\chi\sim 0.4$, where $\Delta=M_\Delta-M_N$ and $\Lambda_\chi\sim m_\rho$ is the chiral scale 
\cite{Pascalutsa:2002pi}.  In this counting, the HB work of ref.~\cite{McGovern:2012ew} is ${\cal O}(e^2\delta^4)$ in the low-energy region. 
The covariant work of ref.~\cite{Lensky:2009uv} is ${\cal O}(e^2\delta^3)$, at which order $\alphae$ and $\betam$ are not free parameters but predicted.
Substantial contributions to one or both come from $\pi N$ and $\pi \Delta$
loops and from $\Delta$-pole graphs, and the final values, which are the result of significant cancellations, are
$\alphae=10.8\pm0.7$ and $\betam=4.0\pm0.7$.  The errors are theory only. Though no fit to data is involved, plots were presented in ref.~\cite{Lensky:2009uv}
to show that the trend of world data up to around 170~MeV is well reproduced in this calculation.  The spin polarisabilities
are also predictions of the theory at this order, and interestingly they are in good agreement with the DR results, including
for instance $\gamma_0=-0.9$; see table~\ref{table:spinpol}.

\begin{table}[tbp]
  \begin{center}
\begin{tabular}{|c |c |c |c |c |}\hline
&$\gamma_{E1E1}$& $\gamma_{M1M1}$& $\gamma_{E1M2}$& $\gamma_{M1E2}$\\ \hline
HB $\delta^3$\cite{Griesshammer:2012we}&  $-5.5$ & $2.1$ & $\hphantom{-}0.5$ & 1.3 \\ \hline
HB $\delta^4$\cite{McGovern:2012ew}&  $-1.1$ & $2.2^*$ & $-0.4$ & 1.9 \\ \hline
Cov. $\delta^3$\cite{Lensky:2014}&  $-3.3$ & $3.0$ & $\hphantom{-}0.2$ &1.1 \\ \hline
DR \cite{Drechsel:2002ar,Griesshammer:2012we}&  $-3.85\pm0.45$ & $2.8\pm0.1$ & $-0.15\pm0.15$ & $2.0\pm0.1$ \\ \hline
\end{tabular}\end{center}
\caption{\label{table:spinpol} Predictions for the spin polarisabilities in three variants of $\chi$PT and in DR, in units of $10^{-4}\fm^4$. 
*This value was fit in ref.~\cite{McGovern:2012ew}; the predicted value would be $\gamma_{M1M1}= 6.4$.} 
\end{table} 

As yet no full calculation has been carried out in the covariant theory at ${\cal O}(e^2\delta^4)$.  The extra graphs required at this order are not only $\pi N$ loop 
graphs with insertions of second-order LECs, namely the proton and neutron anomalous magnetic moments and the $\pi N$ scattering LECs $c_i$, but also 
photon-nucleon seagull graphs with fourth-order LECs which contribute directly to $\alphae$ and $\betam$.  All of these were included in the heavy baryon 
calculations of ref.~\cite{McGovern:2012ew}.  There it was shown that the contribution of the extra loop graphs was quite modest. However the new counter-terms
$\delta\alphae$ and $\delta\betam$ must be fit to Compton scattering data, and seem to be the principal new effect at this order. 
In view of the interest in the apparent discrepancy between DR and \chiEFT\ extractions of $\alphae$ and $\betam$, we consider a partial ${\cal O}(e^2\delta^4)$
covariant result to be of interest, and so we supplement the Lagrangian for proton fields used in \cite{Lensky:2009uv} with the term 
\cite{Fettes:2000gb,Krupina:2013dya}
\begin{equation}
\begin{split}
{\cal L}_{\pi N}^{(4)}= &\pi e^2 \bigl(
 \overline\psi\delta\betam F^{\mu\rho} F^{}_{\mu \rho}\psi\\&-{\textstyle \frac{2}{\MN^2}}(\delta\alphae + \delta\betam)
   (\partial_\mu\overline\psi) F^{\mu\rho} F^{\nu}_{\; \; \rho}\partial_\nu\psi\bigr).
     \label{eq:LpiN4}
   \end{split}
\end{equation}
For a review of the power counting, and of the principles underlying the application of \chiEFT\ to Compton scattering, the reader is referred to 
ref.~\cite{Griesshammer:2012we}.

The database of experimental Compton scattering results for energies below 170~MeV, and its treatment, is the same one as was used in
ref.~\cite{Griesshammer:2012we,McGovern:2012ew}; the following is only the briefest of summaries and a thorough discussion may be found in
ref.~\cite{Griesshammer:2012we}.  The largest single dataset is from TAPS \cite{Olmos:2001},
for which we allow a point-to-point systematic error of 4\% as advocated by Wissmann \cite{Wissmann:2004}; other modern data is
from \cite{Federspiel:1991,Zieger:1992,Hallin:1993,MacGibbon:1995}, and a number of older experiments also contribute some points.
Normalisation uncertainties are incorporated into the $\chi^2$ function in the usual way. As in ref.~\cite{Griesshammer:2012we,McGovern:2012ew}
we  used the data of Hallin \etal\ \cite{Hallin:1993} below 150~MeV only.  

\begin{figure}[!t]
  \begin{center}
    \includegraphics[width=.75\columnwidth]{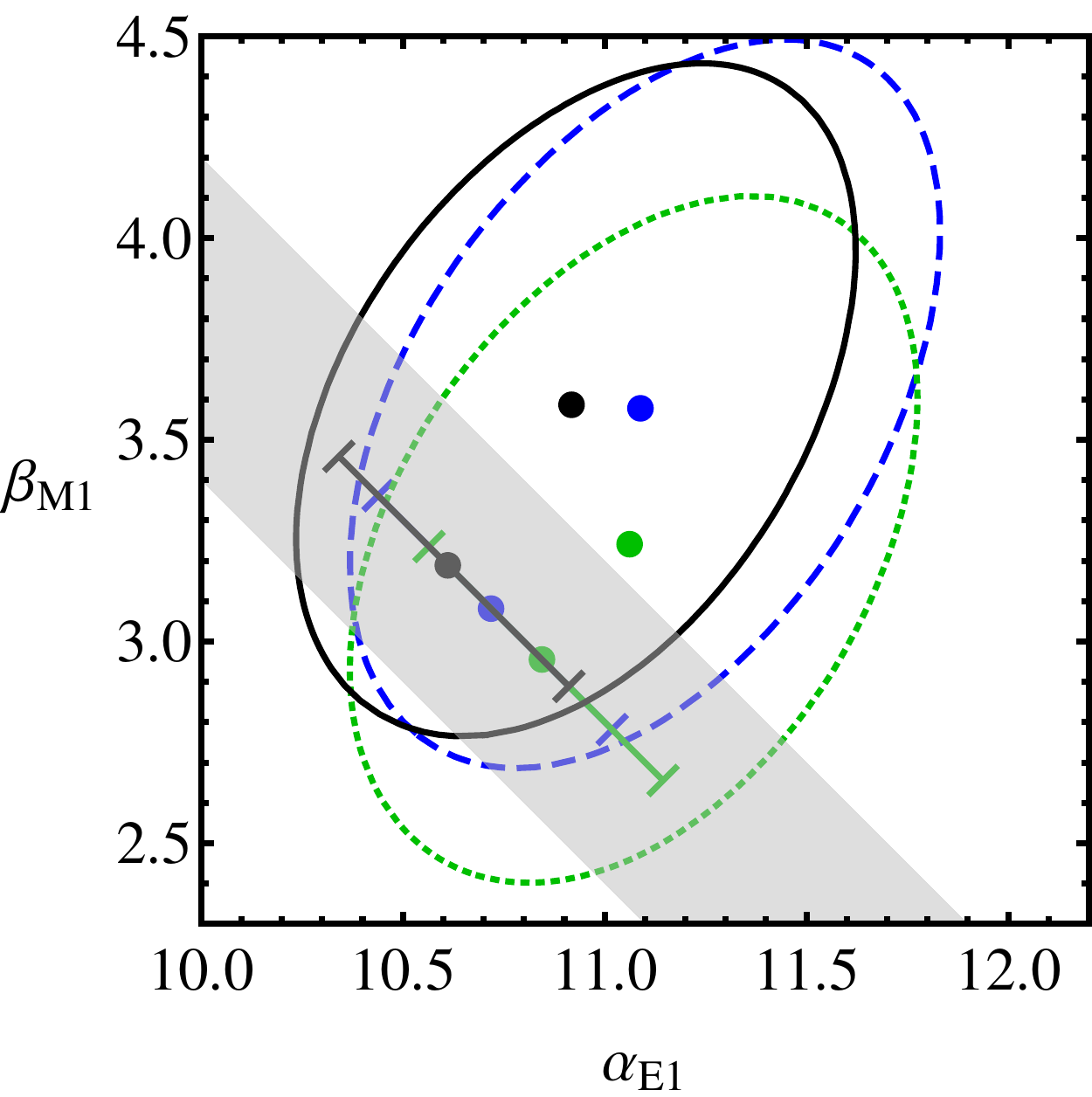}
  \end{center}
 \caption{\label{fig:chi} (Colour online) One sigma contours for the one parameter (Baldin-constrained) and two parameter (unconstrained) fits,
 for three variants of the data set: the solid (black) line is our final result, while the dashed (blue) lines exclude all Hallin data and the dotted (green)
 line uses an upper cut-off of 150~MeV. These are sets (III), (II) and (I) respectively of ref.~\cite{McGovern:2012ew}.
}
\end{figure}

\begin{figure*}[tb]
  \begin{center}
    \includegraphics[width=2\columnwidth]{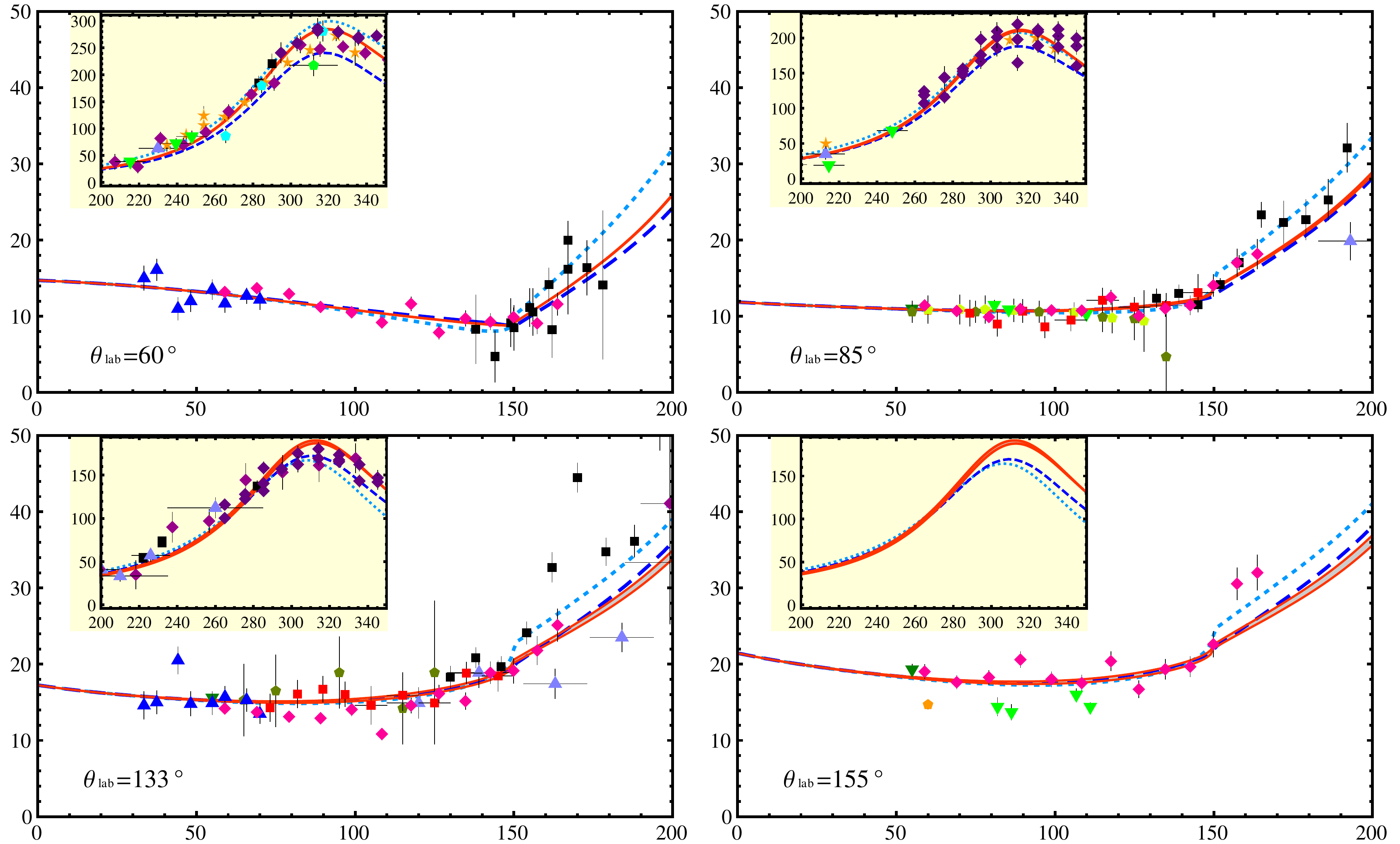}
  \end{center}
 \caption{\label{fig:data} (Colour online) The predictions for the cross section in nb/sr as a function of incoming photon energy in MeV (both in lab frame), 
using the values  from the Baldin-constrained fit
$\alphae=10.6$ and $\betam=3.2$, and with the band showing statistical errors only (solid line, red).  For comparison the results of \HBCPT\ are shown: 
${\cal O}(e^2\delta^3)$ \cite{Griesshammer:2012we} (dotted, light blue) and ${\cal O}(e^2\delta^4)$ \cite{McGovern:2012ew}
(dashed, dark blue). Data points are shown without floating normalisation, and including those points which were excluded from the fit.
The key is given in Table 3.1 of ref.~\cite{Griesshammer:2012we}; in particular purple diamonds are Mainz data \cite{Olmos:2001,Wolf:2001ha};
 red squares, MacGibbon \cite{MacGibbon:1995}; black squares, Hallin \cite{Hallin:1993}; and  blue triangles,  Federspiel \cite{Federspiel:1991}.}
\end{figure*}

We do not fit to any higher-energy data but check by eye that the agreement continues to be good at higher energies (at least as far as the deficiencies
of the data sets allow such a judgement). We take the $\gamma N\Delta$ coupling constants obtained in a fit to photoproduction
data: $g_M=2.97$ and $g_E=-1.0$ \cite{PV06}.  Other parameters are given in ref.~\cite{McGovern:2012ew}.
We present results both with and without imposing the Baldin sum rule 
$\alphae + \betam= 13.8\pm0.4$ \cite{Olmos:2001}, as follows:
\begin{align}
  \alphae&=10.9\pm0.45(\text{stat})\pm0.4(\text{theory})\nonumber\\
  \betam &=3.6\pm0.55(\text{stat})\pm0.4(\text{theory}) 
  \label{eq:4th-fit-NB}
\end{align}
with $\chi^2$ of 111.8 for 135 d.o.f. For the Baldin-constrained fit
we obtain $\alphae - \betam= 7.4\pm0.5(\text{stat})\pm0.4(\text{theory})$, giving
\begin{align}
    \alphae=&10.6 \pm0.25(\text{stat})\pm0.2(\text{Baldin})
    \pm0.4(\text{theory})\nonumber \\
    \betam =&3.2\mp0.25(\text{stat})\mp0.2(\text{Baldin})
    \pm0.4(\text{theory})
  \label{eq:4th-fit-Bal}
\end{align}
with  $\chi^2$ of 112.5 for 136 d.o.f.  The theory errors have been conservatively calculated, based on the shift of $\alphae - \betam$
 from $e^2\delta^3$ to partial $e^2\delta^4$, multiplied by the parameter $\delta\sim0.4$.

These results are completely consistent with one another, as can be seen from Fig.~\ref{fig:chi} which also demonstrates the sensitivity
to variations of the choice of database.  Furthermore they are consistent with the corresponding results in the heavy baryon extraction \cite{McGovern:2012ew},
and in fact the central values of the Baldin-constrained fit are essentially equal to those of that work, namely $\alphae=10.65\pm0.35(\text{stat})$, 
$\betam =3.2\mp0.35(\text{stat})$. 
(The statistical error is  higher than in the current work because an extra parameter, $\gamma_{M1M1}$, was fitted in ref.~\cite{McGovern:2012ew}.)  
It is also interesting to recall that partial and full ${\cal O}(e^2\delta^4)$ HB extractions of $\alphae - \betam$ agree very well~\cite{McGovern:2012ew}.

In Fig.~\ref{fig:data} we show the fit along with a selection of data. The low-energy fit is excellent, and below the photoproduction threshold
the predictions of covariant and heavy-baryon $\chi$PT are largely indistinguishable.  (The similarity of the covariant and HB predictions for the unpolarised
cross sections in this region was already noted in ref.~\cite{Lensky:2012ag}.) The values of $\alphae$ and $\betam$ in all three cases
are extremely close, but the spin polarisabilities are quite disparate.   The  HB ${\cal O}(e^2\delta^3)$ curves give a stronger cusp than
the covariant version, but the HB ${\cal O}(e^2\delta^4)$ is in very good agreement with the covariant calculation up to 200~MeV and beyond.
Though only low-energy data is used in the fit, the good agreement with the Mainz data \cite{Wolf:2001ha} continues into the resonance region, though at most
angles the Delta peak is somewhat too high. It should be noted though that in this region the power counting changes and the EFT calculation is only NLO.
Moreover the HB fits varied the $\gamma N\Delta$ coupling constant $g_M$ whereas
in the present, covariant fit we have used the value obtained in the covariant theory from 
photoproduction  \cite{PV06}, which is around 10\% higher.
The leading dependence of the height of the peak on the coupling constant is $g_M^4$, and a better fit in the resonance region could be obtained by allowing
a modest variation of this parameter with negligible effect on the low-energy fit. 

In summary, we have shown that, in a fit to low-energy Compton scattering data, very similar results are obtained for the electromagnetic 
polarisabilities of the proton, $\alphae$ and $\betam$, whether the covariant or heavy baryon versions of chiral effective field theory are used.  
In particular if the 
Baldin sum-rule constraint is applied, the extracted values of $\alphae-\betam$ are essentially identical. This result is unexpected because some other
predictions of the two versions, notably the spin polarisabilities, are not in good agreement. However these are not the dominant drivers of the 
energy evolution at photon energies comparable to $\mpi$, and the two versions of the theory make very similar predictions 
for the overall cross section. 
It should be noted that this energy dependence, including the cusp at photoproduction threshold generated by chiral loops, is highly non-trivial. 
The excellent fit to data with only one free parameter demonstrates the predictive power of \chiEFT.

It is still to be tested whether the tension 
between the \chiEFT\ extraction and the widely accepted dispersion-relation-based one of 
ref.~\cite{Olmos:2001} is due to the 
larger and more carefully handled dataset used in the chiral extractions, or to some other feature of the predictions of the two theories.  
\add{More unpolarised  data, particularly at energies around $\mpi$ at backward angles, might be needed to resolve the issue. To further 
explore  spin polarisabilities, though, it is clear that polarised scattering measurements will be required.}

\begin{acknowledgments} We are grateful to Daniel Phillips and Harald Grie\ss hammer for discussions regarding aspects of these calculations,
and to Mike Birse and Vladimir Pascalutsa for useful comments on the manuscript.
This work has been supported in part by UK
Science and Technology Facilities Council grants ST/F012047/1, ST/J000159/1
\end{acknowledgments}

  \end{document}